\documentclass[12pt]{spieman}  
\usepackage{amsmath,amsfonts,amssymb}
\usepackage{graphicx}
\usepackage{setspace}
\usepackage{tocloft}
\usepackage{lineno}

\title{On the first quantization and quantum diversity of photons }

\author{Boris Chichkov}
\affil{Leibniz University Hannover, Institute of Quantum Optics, Welfengarten Srt. 1, 30167 Hannover, Germany }

\cftpagenumbersoff{figure}
\cftpagenumbersoff{table} 
\begin{document} 
\maketitle

\begin{abstract}
Quantum theory of photons based on the first quantization technique,  similar to that used by Schr\"odinger in the formulation of quantum mechanics, is considered. First, scalar quantum mechanics of photons operating with the photon wave functions is discussed. Using the first quantization,  the wave equation, the Schr\"odinger-like equations, and the Dirac equation for photons  are derived. Then, vector quantum mechanics of photons is introduced, which defines the electromagnetic vector fields. Using the first quantization, the Maxwell equations for photons in magneto-dielectric medium are obtained. Since the photon electric and magnetic fields satisfy the Maxwell equations, all what is known about the classical optical fields can be directly transferred to 
photons demonstrating their quantum diversity.  Relationships between the scalar and vector quantum mechanics of photons and between the Dirac and Maxwell equations are analyzed. To describe the propagation of photons in dispersive media novel equations are introduced.     
\end{abstract}


{\noindent \footnotesize\textbf{*} Adress correspondence to \linkable{chichkov@iqo.uni-hannover.de} }

\begin{spacing}{1}   

\section{Introduction}

For celebration of 100 years of quantum science, the year 2025 is announced by UN and UNESCO as the international year of quantum science and technology.  One of the most important celebrated fields is quantum photonics which is rapidly developing due to advances in single photon sources\cite{Fei,Ara, Boz, Esm}. Although the concept of photon is widely accepted by the quantum optics community, it can lead to confusions and misunderstandings\cite{Loudon}. Very often, a photon is considered as a quantum of electromagnetic radiation confined in an optical cavity and defined as an elementary excitation of a single cavity mode, characterized by the frequency $\omega$, the wave vector ${\bf k}$, and polarization of the quantized electromagnetic field.  The mode ${\bf k}$ of the radiation field is  analogous to a one-dimensional simple harmonic oscillator with the energy $E_k=(n_k+1/2)\hbar\omega $, where $n_k$ is the number of photons in this mode. This definition corresponds to a monochromatic photon fully delocalized in time. This is probably the reason why Lamb considered photon as a bad concept\cite{Lamb}. In practice, the photon is defined as a tiny energy package of electromagnetic radiation emitted by a single atom, molecule, quantum dot, or other single photon source, which has a certain time duration $\Delta \: t$, defined by the radiative decay probability, and frequency bandwidth $\Delta\omega$ satisfying the Heisenberg uncertainty relation $\Delta\omega \Delta \: t \sim 2\pi$. 
This is the multimode propagating photon with a certain degree of localization in time and in space, in a volume $V>>\lambda^3$, where $\lambda$ is the wavelength corresponding to the maximum in the photon frequency spectrum $a(\omega)$, with $\int a(\omega)d\omega=1$.
Mathematically, one can describe this multimode photon as a superposition of monochromatic photon modes. This is a more general approach to single photon states, including photon wave-packets, which is applied in quantum optics\cite{Cohen, Keller, Boyd}. For a semi-popular view on the nature of photon, a collection of papers edited by C, Roychoudhuri and R. Roy can be recommended\cite{Chan}.

The wave function of a photon is a concept  extending the principles of quantum mechanics to massless particles like photons. Since the photon is associated with the electromagnetic field, its wave function is often expressed in terms of the vector potential $\bf{A}(\bf{r},t)$ or the electric and magnetic field vectors $\bf{E}(\bf{r}, t)$ and $\bf{B}(\bf{r}, t)$. This wave function  is not a scalar function, but vector function, which is responsible for some interpretation difficulties. Discussions of different formulations of photon wave function and photon quantum mechanics  are still ongoing and represent an important hot topic\cite{Sipe, Gers, Kel, Smith, Cug, Barnet, Marg, Hod}.

The aim of this paper is to provide a direct and simple approach to the quantum theory of photons  based on the so-called "first quantization" technique similar to that used by Schr\"{o}dinger in the development of quantum mechanics. First quantization deals with a fixed number of particles and converts classical energy conservation equations into quantum wave equations by replacing classical observables like energy and momentum by operators acting on the wave function.  In second quantization the number of particles is not fixed and the involved interaction fields become quantized through the introduction of creation and annihilation operators. In contrast to "second quantization", the result of first quantization of a photon is not generally known, so I consider it is worthwhile to discuss it in this paper.

\section{First quantization of photons}

Due to the wave-particle duality of matter, we can consider a photon  in a dielectric medium with the refractive index $n$ as 
 a particle with the following relation between the photon energy  $E=\hbar\omega$ and its momentum ${\bf p}=\hbar {\bf k}$ 
\begin{equation}\label{ec}
E=pc/n, 
\end{equation}
where $p=|{\bf p}|=n\hbar\omega/c$ is the Minkowski expression for the photon momentum. This expression is broadly used in photonics, laser physics, and nonlinear optics, and it is fully consistent with the special relativity.\cite{me}  
Eq. (\ref{ec}) allows to formulate quantum mechanics of photons and to derive analogs of the Schr\"{o}dinger and Dirac equations.  Note that Eq.  (\ref{ec}) is fulfilled for all frequency components of the photon spectrum. It can be written in the following form  $Ea(\omega)=p a(\omega)c/n$, where $a(\omega)$ can be canceled. 

Throughout this paper, we use the CGS (centimeter–gram–second) Gaussian system of units, where the refractive index $n=\sqrt{\epsilon\mu}$, and $\epsilon$, $\mu$ are the permittivity and permeability of the optical medium. The advantage of the SGS Gaussian system is that the electric and magnetic fields have the same units (cm$^{-1/2}$g$^{1/2}$ s$^{-1}$) and the speed of light $c$ explicitly enters in all equations.

One of the solutions of the quantum equation for photons that we are looking for should be the plane wave 
\begin{equation}\label{plane}
\Psi ({\bf r}, t)\sim \exp{\left(-i\frac{E}{\hbar} t+i\frac{{\bf p r}}{\hbar} \right)}. 
\end{equation}
Differentiation over time and coordinates allows to define the energy $ \hat E=i\hbar \partial/\partial t$ and momentum ${\bf \hat p}=-i\hbar\partial/\partial {\bf r}=-i\hbar \nabla $ operators as in the standard quantum mechanics. For the plane wave propagating in the opposite ($-{\bf r}$) direction, $+$ sign in the plane wave should be replaced by $-$, and the sign of the corresponding momentum operator will be changed, ${\bf \hat p}=i\hbar \nabla $. Thus, for a photon trapped in a resonator cavity, which can be presented by two counter propagating plane waves, the momentum operator should be defined by ${\bf \hat p}=\mp i\hbar \nabla $. 

To use the energy and momentum operators in Eq.(\ref{ec}), we need to define the operator for absolute value of the photon momentum $p=|{\bf p}|$. There are several options for that, which will be discussed below. Here, to overcome this definition problem, we take the square of Eq. (\ref{ec}) and rewrite it in the following form
\begin{equation}\label{ec2}
E^2={\bf p}^2\left(\frac{c}{n}\right)^2. 
\end{equation}
Now we can use the quantization operators directly, introducing them in Eq. (\ref{ec2}), and write the corresponding quantum equation for a free propagating photon (and a photon trapped in a resonator cavity) in the following form
\begin{equation}\label{we}
\left(\frac{\partial^2}{\partial t^2}-\frac{c^2}{n^2}\Delta\right)\Psi({\bf r},t)=0,
\end{equation}
where $\Delta=\nabla^2$. This is the quantum wave equation for a photon with the scalar photon wave function $\Psi({\bf r},t)$ satisfying the normalization equation $\int |\Psi({\bf r},t)|^2 d^3{\bf r}=1$, where the integration is performed over entire space. 
The $\Psi({\bf r},t)$ function defines the probability to find photon at a given time in a defined spatial region with the volume $V>>\lambda^3$. Note that this equation coincides with the well-known classical wave equation for an optical wave. Photon electric ${\bf E}({\bf r},t)$ and magnetic ${\bf H}({\bf r},t)$ fields also satisfy the wave equation (\ref{we}) and the following normalizations
\begin{equation}
\frac{\epsilon}{8\pi}\int |{\bf E}({\bf r},t)|^2 d^3{\bf r}=\hbar\omega, \quad\quad
\frac{\mu}{8\pi}\int|{\bf H}({\bf r},t)|^2 d^3{\bf r}=\hbar\omega, 
\end{equation}
where $\hbar\omega$ is the photon energy, averaged over the photon spectrum. This allows the following relationships between the electric and magnetic fields and the wave function to be defined 
\begin{equation}
 |{\bf E}({\bf r},t)|=\sqrt{\frac{8\pi}{\epsilon}\hbar\omega}\;|\Psi ({\bf r}, t)|, \quad\quad
|{\bf H}({\bf r},t)|=\sqrt{\frac{8\pi}{\mu}\hbar\omega}\;|\Psi ({\bf r}, t)|.
\end{equation}

If in Eq. (\ref{ec2}) only the momentum is replaced by the operator, we get the Helmholtz equation for a photon 
\begin{equation}\label{H}
\left(\Delta +k^2\right) \Psi({\bf r},\omega)=0,
\end{equation}
with $k=\omega n/c$. This is the wave equation in the frequency domain for the Fourier transform $ \Psi({\bf r},\omega)$ of the wave function defined by  
\begin{equation}\label{F}
 \Psi({\bf r},t) = \int\Psi({\bf r},\omega)\exp( i\omega t)d\omega,
\end{equation}
which represents the eigenvalue problem for the Laplace operator. 
If in Eq. (\ref{ec2}) only the energy is replaced by the operator, we get the following equation for the Fourier transform $\Psi({\bf k},t)$
\begin{equation}\label{H1}
\left(\frac{\partial^2}{\partial t^2}+\omega^2\right)\Psi({\bf k},t)=\left(\frac{\partial}{\partial t}+i\omega\right)\left(\frac{\partial}{\partial t}-i\omega\right)\Psi({\bf k},t)=0,
\end{equation}
defined by 
\begin{equation}\label{F1}
 \Psi({\bf r},t) = \int\Psi({\bf k},t)\exp( i{\bf k} {\bf r})d^3{\bf k},
\end{equation}
which can be proven by inserting this expression into the wave equation (\ref{we}). The formal solution of 
Eq.(\ref{H1}) can be written in the following form
\begin{equation}\label{S}
 \Psi({\bf k},t) = A({\bf k})\exp( \pm i\omega t),
\end{equation}
where $A({\bf k})$ is an arbitrary function satisfying the normalization condition.

Using the slowly varying envelope approximation for the wave equation or paraxial approximation for the Helmholtz equation, further simplifications known from classical optics can be obtained which allow to introduce Gaussian pulse-like photons or Gaussian beam-like photons.

\section{Scalar quantum mechanics of photons}

\subsection{Schr\"odinger like equations for photon}

In scalar quantum mechanics of photons we introduce operators acting on the scalar wave functions and can rewrite Eq (\ref{ec}) in the following form
\begin{equation}\label{ecs}
E=\frac{({\bf k}\cdot{\bf p})}{k}\frac{c}{n}, 
\end{equation}
where $k=|{\bf k}|$. Inserting operators for energy and momentum for a free propagating photon in this equation, we get
\begin{equation}\label{sch}
i\hbar\frac{\partial}{\partial t}\Psi ({\bf r}, t)=\hat H\Psi ({\bf r}, t), \quad \quad \hat H=-i\hbar\frac{c}{n} \frac{({\bf k}\cdot \nabla)}{k}
\end{equation}
This equation is similar to the Schr\"odinger equation with the hamiltonian operator $\hat H$. Choosing the coordinate system with $z$ axis parallel to ${\bf k}$, it can be written in the following form
\begin{equation}\label{sch1}
\left(\frac{\partial}{\partial t}+\frac{c}{n}\frac{\partial}{\partial z}\right)\Psi (z, t)=0.
\end{equation}
This equation describes a free propagating photon along the $z$ axis. For a photon trapped in a resonator (e.g. between two mirrors) ${\bf k}$ vector can be directed parallel and antiparallel to the $z$ axis. For the photon moving in the opposite direction to the $z$ axis, we get 
\begin{equation}\label{sch2}
\left(\frac{\partial}{\partial t}-\frac{c}{n}\frac{\partial}{\partial z}\right)\Psi (z, t)=0.
\end{equation}
For a photon trapped in a resonator cavity, the Hamiltonian operator
\begin{equation}\label{ham}
\hat H=\pm \frac{c}{n} ({\bf e}_k\cdot {\bf \hat p})=\mp i\hbar\frac{c}{n} ({\bf e}_k\cdot \nabla),
\end{equation}
where ${\bf e}_k= {\bf k}/k$, ${\bf \hat p}=-i\hbar \nabla $, and Eqs.(\ref{sch1},\ref{sch2}) have to be simultaneously fulfilled, which again results in the wave equation written here for ${\bf k}$ parallel to the $z$ axis 
\begin{equation}\label{sch3}
\left(\frac{\partial}{\partial t}+\frac{c}{n}\frac{\partial}{\partial z}\right)\left(\frac{\partial}{\partial t}-\frac{c}{n}\frac{\partial}{\partial z}\right)\Psi (z, t)=\left(\frac{\partial^2}{\partial t^2}-\frac{c^2}{n^2}\frac{\partial^2}{\partial z^2}\right)\Psi(z,t)=0.
\end{equation}
One can see that an arbitrary function with the arguments $(t\mp zn/c)$ or $(\omega t\mp kz)$, e.i. $\Psi(t\mp zn/c)$ or $\Psi(\omega t\mp kz)=\Phi(\omega t\mp kz)\exp(-i\omega t \pm ikz)$ satisfy the above equation. The same is also valid for electric and magnetic fields.

\subsection{Pauli spin matrices and Dirac equation for photon}

Photons are bosons with the spin equal 1 and helicity $\pm 1$, defined as the projection of the spin onto the direction of momentum, corresponding to the two possible pure states of circular polarization, right-handed (+1) and left-handed (-1).  To describe such states it is convenient to introduce the Pauli matrices 
\begin{equation}
   \sigma_x=\begin{pmatrix}
       0 & 1\\
       1& 0
   \end{pmatrix} ,\; \sigma_y=\begin{pmatrix}
       0 & -i\\
       i & 0
   \end{pmatrix}, \; \sigma_z=\begin{pmatrix}
       1 & 0\\
       0 & -1
   \end{pmatrix}
\end{equation}
and the Pauli vector $\boldsymbol{\sigma}=\sigma_x{\bf e}_x +\sigma_y{\bf e}_y+\sigma_z{\bf e}_z$, where ${\bf e}_{x,y,z}$ are the unit coordinate vectors. For us the following expressions for the scalar product of the Pauli vector with the photon momentum are important
\begin{equation}
   \boldsymbol{\sigma}\cdot{\bf p}=\begin{pmatrix}
       p_z & p_x-ip_y\\
        p_x+ip_y& -p_z
   \end{pmatrix} ,\quad (\boldsymbol{\sigma}\cdot{\bf p})^2=\begin{pmatrix}
       {\bf p}^2 & 0\\
        0 & {\bf p}^2, 
   \end{pmatrix} = {\bf p}^2 I,
\end{equation}
where $I$ is the $2\times2$ identity matrix. This allows to rewrite Eq. (\ref{ec2}) in the following matrix form
\begin{equation}\label{ec3}
E^2 I=\left(\frac{c}{n}\right)^2(\boldsymbol{\sigma}\cdot{\bf p})^2. 
\end{equation}
Considering $\hat E=i\hbar \partial/\partial t$ and ${\bf \hat p}=-i\hbar \nabla$ as operators acting on scalar functions, Eq. (\ref{ec3}) defines two quantum equations for spinors $\phi$ and $\chi$, representing two polarization components characterized by (+1) and (-1) helicity.
\begin{equation}\label{Pauli1}
\left(\hat E I -\frac{c}{n}(\boldsymbol{\sigma}\cdot {\bf \hat p})\right)\phi=0, \quad \left(\hat E I +\frac{c}{n}(\boldsymbol{\sigma}\cdot {\bf \hat p})\right)\chi=0, 
\end{equation}
where $\phi=\begin{pmatrix}
       \phi_1 \\
       \phi_2
   \end{pmatrix}$ and $\chi=\begin{pmatrix}
       \chi_1 \\
       \chi_2
   \end{pmatrix}$. Similar to the above discussions, in the coordinate system with z axis parallel to 
the wavevector ${\bf k}$, $\phi$ spinor describes the free propagating photon along the $z$ axis  with 
the $\phi_1$ component having helicity $+1$ and $\phi_2$ component having helicity $-1$, satisfying the normalization condition $\int (|\phi_1|^2+|\phi_2|^2)d^3{\bf r}=1$.  The $\chi$ spinor describes the photon propagating in the opposite direction with +1 and -1 helicities of $\chi_1$ and $\chi_2$ components, respectively. 
The corresponding Hamiltonian for the Schr\"odinger like equation is defined by 
\begin{equation}\label{hamp}
 \hat H=\pm \frac{c}{n} (\boldsymbol{\sigma}\cdot {\bf \hat p}), 
\end{equation}
where $\pm$ corresponds to two different propagation directions.  Both $\phi$ and $\chi$ spinors with 4 components describe the photon trapped in a resonator cavity. Introducing the 4-component wave function with $\Psi_i=\phi_i$ and $\Psi_{i+2}=\chi_i$, where $i=1,2$, one can write the Schr\"odinger equation for the photon trapped in the resonator cavity in a $4\times 4$ matrix form
 \begin{equation}\label{schm}
i\hbar\frac{\partial}{\partial t}\Psi ({\bf r}, t)=\hat H\Psi ({\bf r}, t), \quad \quad \Psi=\begin{pmatrix}
       \Psi_1 \\
       \Psi_2 \\
       \Psi_3\\
       \Psi_4
   \end{pmatrix}, \quad \quad
   \hat H=\begin{pmatrix}
       \frac{c}{n} (\boldsymbol{\sigma}\cdot {\bf \hat p})& 0_m \\
      0_m & -\frac{c}{n} (\boldsymbol{\sigma}\cdot {\bf \hat p})
   \end{pmatrix},
\end{equation}  
where $0_m$ is the $2\times2$ matrix of 0.

There is another way to write Eq. (\ref{ec3}) in the quantum form for spinors $\phi$ and $\chi$
\begin{equation}\label{Dirac}
\hat E I\phi -\frac{c}{n}(\boldsymbol{\sigma}\cdot {\bf \hat p})\chi=0, \quad \hat E I\chi -\frac{c}{n}(\boldsymbol{\sigma}\cdot {\bf \hat p})\phi=0, 
\end{equation}
where $\hat E$ and ${\bf \hat p}$ are operators acting on the spinor components. Introducing the 4-component wave function, as it was done above, one can write the Schr\"odinger equation for the photon trapped in an optical cavity in the form similar to Eq. (\ref{schm})  with the Hamiltonian operator defined by 
\begin{equation}\label{DiracH}
   \hat H_{ph}=\begin{pmatrix}
       0_m & \frac{c}{n} (\boldsymbol{\sigma}\cdot {\bf \hat p}) \\
       \frac{c}{n} (\boldsymbol{\sigma}\cdot {\bf \hat p}) & 0_m
   \end{pmatrix},
\end{equation}  
In this form the obtained Schr\"odinger equation corresponds to the Dirac equation, which for electron with the rest mass $m$ in the potential $V$ has the following Hamiltonian
\begin{equation}\label{DiracHe}
   \hat H_{e}=\begin{pmatrix}
       I(V+mc^2) & c (\boldsymbol{\sigma}\cdot {\bf \hat p}) \\
       c (\boldsymbol{\sigma}\cdot {\bf \hat p}) & I(V-mc^2).
   \end{pmatrix},
\end{equation}  
One can observe that for $V=0$, $m=0$ and $n=1$ both Hamiltonian operators coincide. 
Eqs. (\ref{Dirac}) and the Schr\"odinger equation with the Hamiltonian operator (\ref{DiracH}) represent a generalization of the Dirac equation for the photon in magneto-dielectric medium.

\section{Vector quantum mechanics of photons}

In vector quantum mechanics of photons we introduce operators acting on vector fields starting from Eq. (\ref{ec}), $E=pc/n$. For the photon energy, the operator remains the same, $ \hat E=i\hbar \partial/\partial t$, and we need to define an operator for the absolute value of the photon momentum $p=|{\bf p}|=\sqrt{{\bf p}^2}$. If we use the standard expression for the momentum operator ${\bf\hat  p}=-i\hbar\nabla$, we get $\hat p=\hbar \sqrt{-\nabla^2}=i\hbar\sqrt\Delta$, which does not seem useful. We need to find an operator that acting on a vector function produces a vector, like the curl (or rotor) operator. We consider a class of vector functions perpendicular to the momentum of photon, $\boldsymbol{\Psi}({\bf r},t)\perp {\bf p}$, with the scalar product  ${\bf p}\cdot\boldsymbol{\Psi}=0$, corresponding to $\nabla \boldsymbol{\Psi}=\rm{div}\boldsymbol{\Psi}=0$ in the operator form.  For this class of functions,  the operator $\hat p$ satisfying the following condition $\hat p^2\boldsymbol{\Psi}={\bf \hat p}^2\boldsymbol{\Psi}$ can be defined as
$\hat p=\hbar \nabla\times=\hbar{\rm {rot}}= i {\bf \hat p}\times$, where ${\bf \hat p}$ is the standard momentum operator.
One can see that the above condition is satisfied
\begin{equation}
\hat p^2\boldsymbol{\Psi}=\hbar^2\rm {rot\,(rot}\boldsymbol{\Psi})=\hbar^2[\rm {grad\,(div}\boldsymbol{\Psi})-\Delta\boldsymbol{\Psi}]=-\hbar^2\Delta \boldsymbol{\Psi}={\bf \hat p}^2\boldsymbol{\Psi}.
\end{equation}
This allows to derive the Schr\"odinger equation in the vector form
\begin{equation}\label{Vec}
i\hbar\frac{\partial}{\partial t}\boldsymbol{\Psi}({\bf r}, t)=\hat H\boldsymbol{\Psi}({\bf r}, t), \quad\quad \hat H=\frac{c}{n}\hbar \nabla\times=i\frac{c}{n}{\bf \hat p}\times
\end{equation}
with the Hamiltonian operator acting on the vector field. We look for solutions of this equation in the following form $\boldsymbol{\Psi}=\sqrt{\epsilon}{\bf E}+i\sqrt{\mu}{\bf H}$, where ${\bf D}=\epsilon{\bf E}$ with ${\bf E}$ and ${\bf D}$ the electric and displacement fields,  and ${\bf B}=\mu{\bf H}$ with ${\bf H}$ and ${\bf B}$ the magnetic field and the magnetic flux density.
In this form $\boldsymbol{\Psi}$ is known as the Riemann–Silberstein vector or Weber 
vector\cite{Bi, Seb}. Inserting the $\boldsymbol{\Psi}$ vector in Eq. (\ref{Vec}) and separating the real and imaginary parts (all fields including the permittivity $\epsilon$ and permeability $\mu$ are assumed real). we obtain the well-known Maxwell equations in a dielectric medium (without charges and currents)  
\begin{equation}\label{Max}
\frac{\epsilon}{c}\frac{\partial {\bf E}}{\partial t}= \nabla\times {\bf H}, \quad \quad  -\frac{\mu}{c}\frac{\partial {\bf H}}{\partial t}= \nabla\times {\bf E}
\end{equation}
and ${\rm div}{\bf E}={\rm div}{\bf H}=0$. These Maxwell equations are equivalent to the vector Schrödinger equation and can be considered as the basic equations for the vector quantum mechanics of photons defining electromagnetic fields. Scalar quantum mechanics, on the other hand, defines wave functions and probabilities.

\section{Relationships between the scalar and vector quantum mechanics of photons}

Introducing the momentum operators ${\bf \hat p}$, the above Maxwell equations can be written in the following form
\begin{equation}\label{Max}
i\hbar\frac{\partial {\bf D}}{\partial t}= -c ({\bf \hat p}\times {\bf H}), \quad \quad i\hbar\frac{\partial {\bf B}}{\partial t}= c ({\bf \hat p}\times {\bf E}).
\end{equation}
To derive relations between the scalar and vector equations, we perform a scalar multiplication of Eq. (\ref{Max}) with the Pauli vector
\begin{equation}\label{Max1}
i\hbar \epsilon\frac{\partial {(\boldsymbol{\sigma}\cdot\bf E})}{\partial t}= -c \boldsymbol{\sigma}\cdot ({\bf \hat p}\times {\bf H}), \quad \quad i\hbar \mu\frac{\partial (\boldsymbol{\sigma}\cdot{\bf H})}{\partial t}= c \boldsymbol{\sigma}\cdot ({\bf \hat p}\times {\bf E}).
\end{equation}
Using the following relation between the Pauli and two other arbitrary vectors ${\bf a}$ and ${\bf b}$
\begin{equation}\label{Paulivector}
i\boldsymbol{\sigma}\cdot ({\bf a}\times {\bf b})=(\boldsymbol{\sigma}\cdot {\bf a})(\boldsymbol{\sigma}\cdot {\bf b})-({\bf a}\cdot {\bf b}),
\end{equation}
and applying it to ${\bf \hat p}$, {\bf H} and ${\bf \hat p}$, {\bf E} vector couples, taking into account that ${\bf \hat p}\cdot {\bf H}={\bf \hat p}\cdot {\bf E}=0$, we get
\begin{equation}\label{MaxD}
\hbar \epsilon\frac{\partial {(\boldsymbol{\sigma}\cdot\bf E})}{\partial t}= c (\boldsymbol{\sigma}\cdot {\bf\hat p})(\boldsymbol{\sigma}\cdot {\bf H}), \quad \quad \hbar \mu\frac{\partial {(\boldsymbol{\sigma}\cdot\bf H})}{\partial t}= -c (\boldsymbol{\sigma}\cdot {\bf \hat p})(\boldsymbol{\sigma}\cdot {\bf E}).
\end{equation}
One can see that the above equations, derived from the Maxwell equations, and the photon Dirac equations for spinors,  Eq. (\ref{Dirac}) 
\begin{equation}\label{Diracr}
i\hbar\frac{\partial}{\partial t}\phi =\frac{c}{n}(\boldsymbol{\sigma}\cdot {\bf \hat p})\chi, \quad\quad i\hbar\frac{\partial}{\partial t}\chi =\frac{c}{n}(\boldsymbol{\sigma}\cdot {\bf \hat p})\phi
\end{equation}
coincide when the following relations are fulfilled
\begin{equation}
\phi=\sqrt{\frac{\epsilon}{8\pi\hbar\omega}}\;(\boldsymbol{\sigma \cdot {\bf E})}, \quad\quad     \chi=i\sqrt{\frac{\mu}{8\pi\hbar\omega}}\;(\boldsymbol{\sigma \cdot {\bf H})}.
\end{equation}
Choosing the $z$ axis along the photon momentum corresponding to $E_z=0$, the above relations can be written in a more explicit form corresponding to two possibilities
\begin{eqnarray}
\begin{pmatrix}
       \phi_1 \\
       \phi_2
   \end{pmatrix}=\sqrt{\frac{\epsilon}{8\pi\hbar\omega}}\begin{pmatrix}
       0 \\
       E_x+iE_y
   \end{pmatrix}; \quad\quad \begin{pmatrix}
       \chi_1 \\
       \chi_2
   \end{pmatrix}=i\sqrt{\frac{\mu}{8\pi\hbar\omega}}\begin{pmatrix}
       0 \\
       H_x+iH_y
   \end{pmatrix},  \\    \begin{pmatrix}
       \phi_1 \\
       \phi_2
   \end{pmatrix}=\sqrt{\frac{\epsilon}{8\pi\hbar\omega}}\begin{pmatrix}
       E_x-iE_y \\
       0
   \end{pmatrix}; \quad\quad \begin{pmatrix}
       \chi_1 \\
       \chi_2
   \end{pmatrix}=i\sqrt{\frac{\mu}{8\pi\hbar\omega}}\begin{pmatrix}
       H_x-iH_y\\
       0
   \end{pmatrix},
\end{eqnarray}
describing pure photon states with two different circular polarizations in a resonator cavity. 

\section{On the quantum diversity of photons}

It is not surprising that the first quantization  (quantum mechanics) of photons  results in the Maxwell equations. Weak electromagnetic fields, even at the single photon level, should satisfy the Maxwell equations. 
Therefore, all what is known about classical optical fields can be directly transferred to the quantum states of photons due to the first quantization procedure. We can speak about 
multimode Gaussian beam-like and Gassian pulse-like photons; Bessel beam-like photons; linearly and elliptically polarized photons; frequency chirped photons;   coherent, incoherent, and partially coherent photons; etc.

In the quantum field theory, corresponding to the second quantization, the electromagnetic field is described by the field creation and annihilation operators. The electric and magnetic field components act like canonically conjugate variables, analogous to the position and momentum operators in quantum mechanics. This leads to the quantum fluctuations and  quantum uncertainty for the electric and magnetic fields. Description of these effects and the second quantization of photons are beyond the scope of this paper.   

\section{Quantum mechanics of photons in a dispersive medium}

In a dispersive medium photons propagate with the group velocity $v_g=d\omega/dk=c/n_g$, 
where $n_g=n+\omega \: dn/d\omega$ is the refractive index for the group velocity (the {\it group index}) calculated at the carrier frequency. In this case the relation between the energy and photon momentum is determined by $E=\hbar\omega=pc/n_g$, which is consistent with the special relativity\cite{me}. Here we consider a medium, where the group velocity dispersion  during the propagation of photons, can be neglected. In this case one can neglect effects of temporal stretching and/or compression of photons. This corresponds to the so-called first approximation of the dispersion theory.  To simplify discussions we also assume the optical medium with $\mu=1$.  

Using the first quantization procedure, we can write the wave equation for the photon in such medium 
\begin{equation}\label{wed}
\left(\frac{\partial^2}{\partial t^2}-\frac{c^2}{n_g^2}\Delta\right)\Psi({\bf r},t)=0.
\end{equation}
Here $n$ is replaced  by $n_g$. The corresponding Schr\"odinger and Dirac equations we get by the same replacement.
The Schr\"odinger equation in the vector form
\begin{equation}\label{Vecd}
i\hbar\frac{\partial}{\partial t}\boldsymbol{\Psi}({\bf r}, t)=\hat H\boldsymbol{\Psi}({\bf r}, t), \quad\quad \hat H=\frac{c}{n_g}\hbar \nabla\times=i\frac{c}{n_g}{\bf \hat p}\times
\end{equation}
after the substitution $\boldsymbol{\Psi}=n_g{\bf E} +i{\bf H}$ results in the Maxwell equations corresponding to the first approximation of the dispersion theory
\begin{equation}\label{Maxd}
\frac{n_g^2}{c}\frac{\partial {\bf E}}{\partial t}= \nabla\times {\bf H}, \quad \quad  -\frac{1}{c}\frac{\partial {\bf H}}{\partial t}= \nabla\times {\bf E},.
\end{equation}
with the following expressions for the photon energy density $U$ and intensity $I$ 
\begin{equation}
    U=\frac{n_g^2}{8\pi}E_0^2=\frac{1}{8\pi}H_0^2, \quad\quad I=Uv_g=\frac{cn_g}{8\pi}E_0^2,
\end{equation}
where $E_0$ and $H_0$ are the field amplitudes. Recall that the generally accepted expressions for the energy density and intensity in the dispersive medium are\cite{Landau} 
\begin{equation}
    U=\frac{1}{16\pi}\left[\frac{d(\omega\epsilon)}{d\omega}E_0^2+\epsilon H_0^2\right]=\frac{nn_g}{8\pi}E_0^2=\frac{n_g}{8\pi n}H_0^2, \quad\quad I=Uv_g=\frac{cn}{8\pi}E_0^2, 
\end{equation}
where $\epsilon =n^2$. The expression for electric energy density was obtained by Taylor expansion up to the first order of $dn/d\omega$ and coincides up to the first order with the expression obtained in this paper
\begin{equation}
    U_E=\frac{1}{16\pi}\frac{d(\omega\epsilon)}{d\omega}E_0^2= \frac{1}{16\pi}\left(n^2+2n\omega\frac{dn}{d\omega}\right)E_0^2\simeq\frac{n_g^2}{16\pi}E_0^2.
\end{equation}
In spite that all equations derived in this section have an approximate character, they can be useful for modeling light propagation in dispersive media. 

\section{Conclusions}
Scalar and vector quantum theories of photons based on the first quantization technique,  similar to that used by Schr\"odinger in the formulation of quantum mechanics, have been introduced. The scalar quantum mechanics defines the wave functions and the vector quantum mechanics defines the photon  electromagnetic fields. Using the first quantization technique,  the Wave equation, the Schr\"odinger-like equations, the Dirac equation, and the Maxwell equations for photons have been derived. 
 Since the  electric and magnetic fields of quantum photons and classical optical fields satisfy the Maxwell equations, all what is known from classical optics can be directly transferred to 
photons demonstrating their quantum diversity.  Relationships between the scalar and vector quantum mechanics of photons, i.e. relationships between the wave functions of photons and electromagnetic fields, have been analyzed. Propagation of photons in dispersive media has been considered.

\subsection* {Acknowledgments}
I acknowledge financial support from the  DFG, German 
Research Foundation under Germany’s Excellence Strategy within
the Cluster of Excellence PhoenixD (EXC 2122, Project ID
390833453) and the Cluster of Excellence QuantumFrontiers
(EXC 2123, Project ID 390837967).  I thank Dr. A.  Evlyukhin for critical discussions.


\end{spacing}
\end{document}